\documentclass[twocolumn,showpacs,preprintnumbers,amsmath,amssymb,prc]{revtex4}
\usepackage{graphicx}
\usepackage{dcolumn}
\usepackage{bm}
\begin{document}
\title{Effect of liquid drop model parameters on nuclear liquid gas phase transition}
\author{G. Chaudhuri$^{1,2}$, S. Mallik$^{1}$}
\affiliation{$^1$Physics Group, Variable Energy Cyclotron Centre, 1/AF Bidhan Nagar, Kolkata 700064, India\\
$^2$Homi Bhabha National Institute, Training School Complex, Anushakti Nagar, Mumbai 400085, India}

\begin{abstract}
The phenomenon of liquid-gas phase transition occurring in heavy ion collisions at intermediate energies is a subject of contemporary interest. In statistical models of fragmentation, the liquid drop model is  generally used to calculate the ground state binding energies of the fragments. It is  well known that the surface and symmetry energy of the hot fragments at the low density freeze out can be considerably modified. In addition to this, the level density parameter also has a wide variation.  The effect of variation of these parameters  is studied on  fragmentation observables which are related to the nuclear liquid gas phase transition. The canonical thermodynamical model which has been very successful in describing the phenomenon of fragmentation is used for the study. The shift in transition temperature owing to the variation in liquid drop model parameters has  been examined.
\end{abstract}
\maketitle
\section{Introduction}
The liquid gas phase transition at intermediate energy nuclear reactions is a well studied phenomenon \cite{Siemens,Gross_phase_transition,Bondorf1,Dasgupta_Phase_transition,Chomaz,Borderie2,Borderie}. Different theoretical models, both statistical and dynamical have confirmed the transition from liquid to gaseous phase as the  excited nuclear system fragments \cite{Gross_phase_transition,Dasgupta_Phase_transition,Chomaz,Borderie2,Borderie,Mallik10,Mallik13}.
This transition is observed in the temperature range of 5 to 6 MeV. The Bethe-Weizsacker mass formula which is commonly referred to as liquid drop model\cite{Bethe,Weizsacker} has successfully explained different ground state properties of the nucleus and is widely used to calculate the binding energy of medium to heavy mass nuclei at zero temperature and normal nuclear density. This has been successfully implemented in statistical models like Canonical Thermodynamical Model(CTM) \cite{Das}, the Statistical Multifragmentation Model (SMM) \cite{Bondorf1} and others in order to throw light on the nuclear multifragmentation process.  Excellent  fits of experimental masses with high level of accuracy for ground state masses at normal density are available \cite{Wang,Mavrodiev,Bingham}. The process of nuclear multifragmentation however occurs at sub saturation density and at higher excitation energies. The density and temperature dependence of the surface and symmetry energy is not incorporated in the simple binding energy formula used in the liquid drop model and hence different observables calculated using this in the statistical models might not be fully reliable. The density and/or temperature dependence of nuclear surface and symmetry energy also plays an important role in areas of astrophysical interest such as the study of supernova explosions and the properties of neutron stars etc. It also has significant influence in deciding the structure of neutron-rich and neutron-deficient nuclei which can be and are formed in fragmentation reactions. In this work we would focus on observables like mass distribution and total multiplicity which are related to the nuclear liquid gas phase transition.  The pertinent question one can ask is that how is the phenomenon of  phase transition dependent on the liquid drop model parameters which dictates the fragmentation pattern. Is the transition temperature sensitive to the parameters of the liquid drop model? These questions motivated us to reexamine the nuclear phase transition process in the framework of the liquid drop model.\\
\indent
One of the important term determining the path of fragmentation is the surface tension or the surface energy coefficient. The competition between the surface term and the excitation energy term of the fragments ultimately dictates the fragmentation pattern, or in other words the liquid gas phase transition. The surface term for obvious reasons favours larger fragments while the other term promotes breaking up into smaller pieces. This establishes the direct connection of the liquid drop model parameters with the phenomenon of phase transition and motivates us to examine in details the effect of these parameters on the later.  The effect of the surface and asymmetry term of the liquid drop model on isotopic scaling and mean neutron to proton ratios has been studied in detail in the framework of the statistical multifragmentation (SMM) model\cite{Ogul2,Ogul1}. But the effect of the same on the nuclear liquid-gas phase transition has not been examined so far and this work was motivated by that. The effect of the temperature dependence of the surface term is also examined in order to study its influence if any on the phase transition. This study is expected to throw light on the relative importance of the liquid drop model parameters while characterizing the liquid gas phase transition. The results from this study can lead to more refined calculation of those parameters of the liquid drop model term which dominates in deciding the phase transition in order to have detailed knowledge about the nature of the transition and its characteristics. One can have more sophisticated models for determining the temperature and density dependence of these relevant terms. In addition to this the effect of the variation of the level density parameter which governs the excitation energy term (from Fermi gas Model \cite{Bohr} has also been examined w.r.t nuclear liquid gas phase transition. In the results presented in this work, we have used the temperature derivative of multiplicity \cite{Mallik16,Mallik19,Mallik20,Lin,Wada} in order to pinpoint the transition temperature as total multiplicity can be easily measured both theoretically as well as experimentally. The multiplicity derivative has already been established both theoretically \cite{Mallik16,Mallik19,Lin} and experimentally \cite{Wada} as a convincing signature of nuclear liquid gas phase transition. In the next section we give a brief description of our model followed by the results section  and finally the summary.\\
\\
\indent
\section{Model description}
 We have used the canonical thermodynamical model(CTM) \cite{Das} in order to study the fragmentation of nuclei. In such models of nuclear disassembly it is assumed that statistical equilibrium is attained at freeze out stage and population of different channels of disintegration is solely decided by statistical weights in the available phase space. The calculation is done for a fixed system size, freeze out volume and temperature. In a canonical model \cite{Das}, the partitioning is done such that all partitions have the correct $A_{0},Z_{0}$ (equivalently $N_{0},Z_{0}$). The canonical partition function is given by
\begin{eqnarray}
Q_{N_{0},Z_{0}} & = & \sum\prod\frac{\omega_{N,Z}^{n_{N,Z}}}{n_{N,Z}!}
\end{eqnarray}
where the sum is over all possible channels of break-up (the number of such channels is enormous) satisfying $N_{0}=\sum N\times n_{N,Z}$ and $Z_{0}=\sum Z\times n_{N,Z}$; $\omega_{N,Z}$ is the partition function of the composite with $N$ neutrons \& $Z$ protons and $n_{NZ}$ is its multiplicity. The partition function $Q_{N_{0},Z_{0}}$ is calculated using a recursion relation \cite{Chase}. From Eq. (1), the average number of composites with $N$ neutrons and $Z$ protons is given by
\begin{eqnarray}
\langle n_{N,Z}\rangle & = & \omega_{N,Z}\frac{Q_{N_{0}-N,Z_{0}-Z}}{Q_{N_{0},Z_{0}}}
\end{eqnarray}
\indent
The partition function of a composite having $N$ neutrons and $Z$ protons is a product of two parts: one is due to the the translational motion and the other is the intrinsic partition function of the composite:
\begin{eqnarray}
\omega_{N,Z}=\frac{V}{h^{3}}(2\pi mT)^{3/2}A^{3/2}\times z_{N,Z}(int)
\end{eqnarray}
where $V$ is the volume available for translational motion. Note that $V$ will be less than $V_{f}$, the volume to which the system has expanded at break up (freeze-out volume). We use $V=V_{f}-V_{0}$ , where $V_{0}$ is the normal volume of nucleus with $Z_{0}$ protons and $N_{0}$ neutrons. In this work the freeze-out volume is kept constant at $6V_0$. For nuclei in isolation, the internal partition function is given by $z_{N,Z}(int)=\exp[-\beta F(N,Z)]$ where $F=E-TS$. For mass number $A \geq$5, we use the liquid-drop formula for calculating the binding energy and the contribution for excited
states is taken from the Fermi-gas model.\\
We now list the properties of the composites used in this work.  The proton and the neutron are fundamental building blocks, thus $z_{1,0}(int)=z_{0,1}(int)=2$, where 2 takes care of the spin degeneracy.  For deuteron, triton, $^3$He and $^4$He we use $z_{N,Z}(int)=(2s_{N,Z}+1)\exp[-\beta E_{N,Z}(gr)]$ where $\beta=1/T, E_{N,Z}(gr)$ is the ground-state energy of the composite and $(2s_{N,Z}+1)$ is the experimental spin degeneracy of the ground state.  Excited states for these very-low-mass nuclei are not included.
For mass number $A\ge5$ we use the liquid-drop formula.  For nuclei in isolation, this reads\\
\begin{eqnarray}
z_{N,Z}(int)
&=&\exp\frac{1}{T}\bigg{[}W_0A-a_s(T)A^{2/3}-a^{*}_c\frac{Z^2}{A^{1/3}}\nonumber\\
&&-C_{sym}\frac{(N-Z)^2}{A}+\frac{T^2A}{\epsilon_0}\bigg{]}
\end{eqnarray}

The expression includes the volume energy [$W_0=15.8$ MeV], the temperature dependent surface energy
[$a_s(T)=a_{s0}\{(T_{c}^2-T^2)/(T_{c}^2+T^2)\}^{5/4}$ with $a_{s0}=18.0$ MeV and $T_{c}=18.0$ MeV], the Coulomb energy [$a^{*}_c=0.31a_{c}$ with $a_{c}=0.72$ MeV and Wigner-Seitz correction factor 0.31 \cite{Bondorf1}] and the symmetry energy ($C_{sym}=23.5$ MeV).  The term $\frac{T^2A}{\epsilon_0}$ ($\epsilon_{0}=16.0$ MeV) represents contribution from excited states since the composites are at a non-zero temperature. The different coefficients in the liquid drop model  and the Fermi Gas Model are fixed empirically and hence there is some uncertainty in their magnitude. In our calculation we will try to examine how the variation of these different parameters can affect the liquid gas phase transition, more specifically the transition temperature.
We have investigated the effect of the surface energy coefficient $\sigma(T)$, the symmetry energy coefficient $C_{sym}$ and the factor  ${\epsilon_0}$ which is connected to the level density parameter $a$ by $a=A/\epsilon_0$ . This will help us to conclude about the sensitivity of these parameters in determining the nuclear liquid gas phase transition. We have tested the effect on transition temperature of conversion from liquid to gas phase and have used the multiplicity derivative $dM/dT$ (M being the total multiplicity) with respect to temperature T as the observable. In a recent work \cite{Mallik19}, we have shown that the peak of the multiplicity derivative and that of the specific heat occurs at the same temperature which has been identified as the transition temperature.\\

\begin{figure}[b]
\begin{center}
\includegraphics[width=5.5cm,keepaspectratio=true]{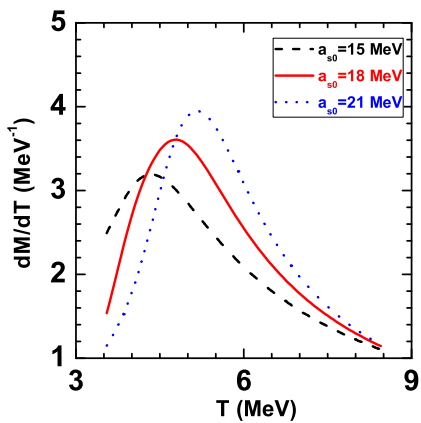}
\caption{Dependence of multiplicity derivative on surface energy co-efficient ($a_{s0}$).}
\end{center}
\end{figure}
\begin{figure}[t]
\begin{center}
\includegraphics[width=0.95\columnwidth]{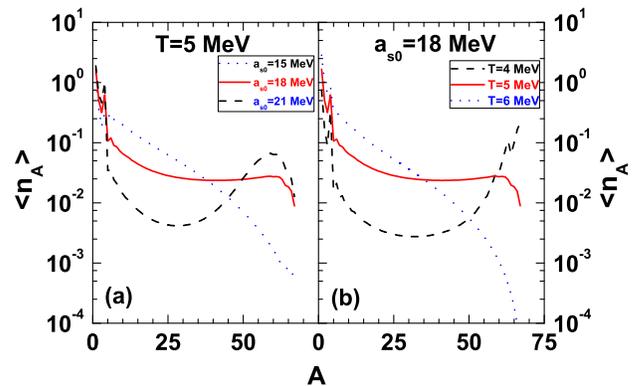}
\caption{Mass distribution (a) for three surface energy at fixed temperature and (b) at three different temperature for fixed surface energy.}
\end{center}
\end{figure}
\begin{figure}[b]
\begin{center}
\includegraphics[width=5.5cm,keepaspectratio=true]{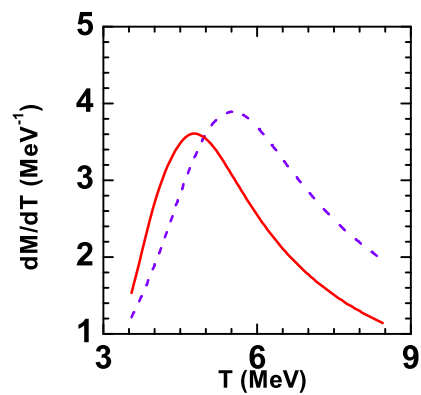}
\caption{Variation of multiplicity derivative with temperature for temperature dependent and independent (violet dashed line) surface energy co-efficient.}
\end{center}
\end{figure}
\begin{figure}[t]
\begin{center}
\includegraphics[width=5.5cm,keepaspectratio=true]{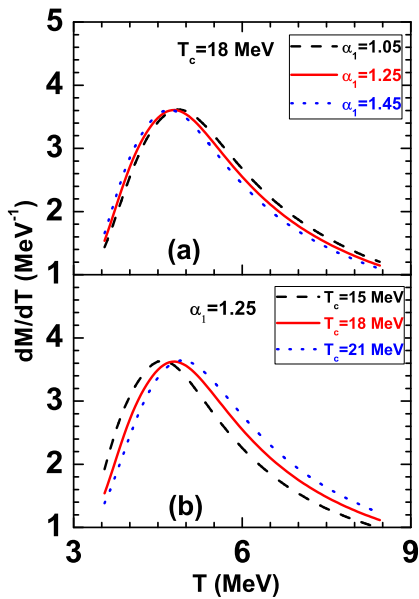}
\caption{Dependence of multiplicity derivative on (a) $\alpha_1$ and (b) critical temperature.}
\end{center}
\end{figure}
\begin{figure}[b]
\begin{center}
\includegraphics[width=\columnwidth,keepaspectratio=true]{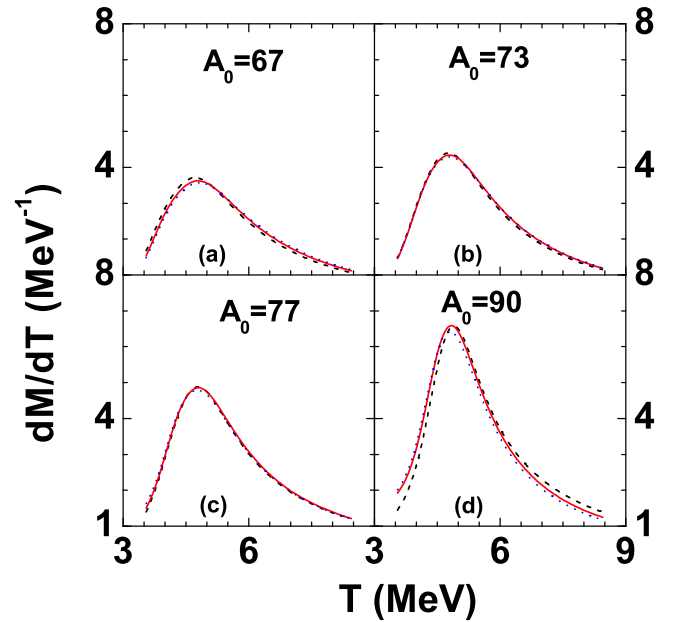}
\caption{Dependence of multiplicity derivative on symmetry energy co-efficient ($a_{sym}$) for four fragmenting sources of same atomic number $Z_0=$32 but different mass number (a) $A_0$=67, (b) 73, (c) 77 and (d) 90. For each case, calculation is done for $a_{sym}=$ 15 MeV(black dashed line), 23.5 MeV (red solid line) and 30 MeV (blue dotted line).}
\end{center}
\end{figure}
\begin{figure}[t]
\begin{center}
\includegraphics[width=5.5cm,keepaspectratio=true]{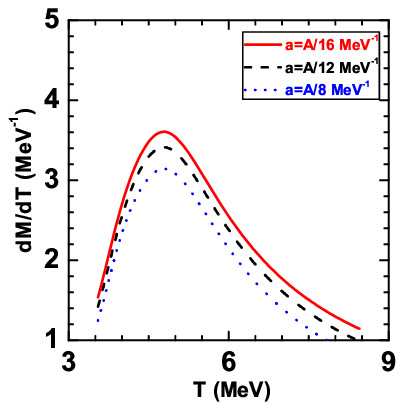}
\caption{Dependence of multiplicity derivative on level density parameter.}
\end{center}
\end{figure}
\indent
\section{Results}
We consider the disintegration of a system of mass number $A_0 =67$  and proton number $Z_0=32$ which is expected to be formed from the central collision of $^{58}Ni$ with $^{9}Be$ without considering pre-equilibrium emission.
The surface energy term  of the liquid drop model is expected to have a  significant role in deciding the phase transition. In order to examine this we have  first calculated  the derivative of total multiplicity as a function of temperature for three different values of the surface energy coefficient keeping all other parameters fixed. This is displayed in Fig. 1 which shows that the peak in the distribution shifts to the right as one increases the surface energy coefficient. This is quite justified  as the surface term will try to hold the nucleus together and hence its increase implies more energy(or temperature) is required for the phase transition from liquid to gas. This explains the shift in transition temperature to the right and the magnitude of shift is about 2 MeV for change in surface coefficient from 15 to 21 MeV. This is quite a significant shift and is expected to affect the transition in a profound manner.  This interesting aspect further motivated us to probe deeper into it and calculate the mass distribution at these different values of the surface coefficient at a fixed temperature. Mass  distribution is a well studied observable which has been experimentally measured in different laboratories across the world and can clearly distinguish between different phases. This is shown in Fig.2(a) and aptly confirms our conclusion that at higher values of the surface  energy coefficient the system is in a coexistence phase and the mass distribution resembles a typical 'U' shape as it should be. With the decrease in the value of $a_s$, the system slowly converts to gaseous phase resulting in disappearance of the peak on the liquid(right) side.  In fact surface energy plays a role equivalent to excitation  energy (or temperature) in dictating the nuclear liquid gas phase transition as will be evident from the figures 2(a) and 2(b). The next figure 2(b) shows the change in mass distribution for a fixed surface energy coefficient as we change the  temperature or the excitation energy. The change in mass distribution of the fragments as we change the temperature(keeping surface energy fixed) is exactly similar to the change as we change the surface energy (keeping temperature fixed). A small change in the surface energy coefficient leads to some major change in the mass distribution as is evident from Fig. 2(a). This explains the magnitude of shift of transition temperature as observed in Fig(1). The exact equivalence of these two figures throws light on the equivalent roles of surface energy and temperature in dictating the phase transition of the nuclear system. The effect of the increase in excitation energy or temperature is equivalent to that of the decrease in the surface energy coefficient.\\
\indent
Having established the importance of surface energy coefficient, it seems mandatory to probe further deep into it and investigate the effect of its temperature dependence on liquid gas phase transition as the fragments are excited. First we would like to show the effect with and without the temperature dependence and the  effect of this on the transition temperature is displayed in Fig. 3. There is a pronounced shift of about 1 MeV in the transition temperature and without using the temperature dependence term the peak shifts to the right implying that the system requires more energy for the transition. This establishes the importance of the temperature dependent term and hence further inspired us to look for the appropriate nature of the temperature dependent term for finite nuclei in the relevant temperature range. Since surface energy term is crucial in fixing the phase transition parameters, hence proper evaluation of its temperature dependence is extremely important in order to have better knowledge about the transition temperature.\\
\indent
In the literature, there has been certain debate about the nature of temperature dependence but the standard prescription in use is the form ${g[T,T_c]}^{\alpha_1}$ where $T_c$ is the critical temperature for finite nuclei and $g(T,T_c)=[T_c^2-T^2]/[T_c^2-T^2]$. While the functional form has been agreed upon to be like this,  there has been some argument as far as the value of ${\alpha_1}$ is concerned. This can assume different values depending on whether its semi infinite nuclear matter or finite nuclei and also the relevant temperature range. The overall variation ranges from  1.05 to 1.45 \cite{Agarwal} depending on the application.  We have investigated the effect of the variation of this on transition temperature once again using the  multiplicity derivative as the signal. It is seen from  Fig4(a) that this parameter has insignificant effect on the phase transition and hence one can continue using the value $1.25$ which has been the usual practice in the Canonical Thermodynamical Model(CTM)\cite{Das}.
In this connection we have also studied the effect of variation of the critical temperature $T_c$ on the nuclear liquid gas phase transition and the results are shown in Fig. 4(b). We have used three values of $T_c$ keeping $\alpha_{1}$ fixed and the results show that there is a small dependence on the value of $T_c$; the transition temperature shifts slightly to the right as the critical temperature is increased but the shift is small.  \\
\indent
The study of the  density and temperature of nuclear symmetry energy is a contemporary topic of research in the domain of nuclear physics as well as nuclear astrophysics \cite{Brown,Steiner,Lattimer2,Danielewicz_science}.  The exact value of this coefficient in the liquid drop model is highly debatable and wide variation has been found in the literature as far as applications in finite nuclei at finite temperature and sub saturation density is concerned. This motivated us to check its impact on phase transition  observables. Values ranging from $15$ to $30$ has been used by different researchers and hence we have checked them accordingly.  It is seen from Fig. 5 that the asymmetry part has very less or almost no effect on  the process of phase transition and hence one can safely use the value $23.5$ for this parameter as is used in the liquid drop model. We have used different isotopes with varying degree of neutron richness ranging from $A_0$ = 67 to 90 in order to confirm our result and the figures establish that the conclusion remains the same for very asymmetric systems far from stability.\\
\indent
The effect of the level density parameter in the excitation energy term is also investigated. This parameter ${\epsilon_0}$ which is related to the level density parameter $a$ ($a=A/ \epsilon_0$) is widely used in heavy ion collisions. It can  vary from 8 to 16 and hence we found it appropriate to examine its effect on phase transition properties. Here also we have used the observable of multiplicity derivative  which has been recently established as a measurable signature for phase transition. It is evident from the calculation and hence the Fig(6)   that this parameter has almost negligible effect on phase transition and we can continue to use the value 16 without having much effect on the study of nuclear liquid gas phase transition.  One can use sophisticated formula for this but there is negligible change for a wide range of the level density parameter. There is  absolutely no shift in the transition temperature with variation in the level density parameter; only the magnitude of the multiplicity derivative changes slightly as is seen from the figure. Similar results like this have also been observed when specific heat was used as an observable instead of multiplicity derivative which further confirms our conclusions. Those results are not shown here for the sake of brevity. More sophisticated formulas \cite{Toke} for level density parameter are available in the literature for evaluation of the level density parameter which are extensively used in the fission studies. These include surface term with deformation dependence which is not of much significance in case of multifragmentation of excited nuclei since the fragments are assumed to be spherical.\\
\\
\indent
\section{Summary}  We have examined the effect of the different parameters used in the liquid drop model and also that in the Fermi gas model on the characteristics of nuclear liquid gas phase transition. More specifically the surface energy coefficient along with its temperature dependence was investigated. The results show that the surface term has a huge impact on the transition temperature and  a small variation can lead to a considerable change.  On the contrary, neither the symmetry energy nor the level density parameter has any significant role in dictating the parameters of the phase transition. This study  thus establishes that it is the surface energy term of the liquid drop model which needs to be determined with more precision using microscopic calculation for better understanding of the phase transition process.

\end{document}